\documentclass[lettersize,journal]{IEEEtran}
\usepackage{hyperref}
\usepackage{amsmath,amsfonts}
\usepackage{amssymb}
\usepackage{algorithmic}
\usepackage{algorithm}
\usepackage{array}
\usepackage{gensymb}
\usepackage{textcomp}
\usepackage{stfloats}
\usepackage{url}
\usepackage{verbatim}
\usepackage{graphicx}
\usepackage{subcaption}
\usepackage{cite}

\newcommand{\argmax}{\mathop{\mathrm{argmax}}} 
\DeclareMathOperator{\vect}{\mathrm{vec}}
\DeclareMathOperator{\abs}{\mathrm{abs}}

\begin{document}
\bstctlcite{IEEEexample:BSTcontrol}
\title{EM-Based Channel Estimation for mMIMO LEO SATCOM Under Imperfect Doppler Compensation}

\author{Abdollah~Masoud~Darya,~\IEEEmembership{Graduate~Student~Member,~IEEE,}
        and~Saeed~Abdallah,~\IEEEmembership{Member,~IEEE}
\thanks{\copyright~2025 IEEE. Personal use of this material is permitted.  Permission from IEEE must be obtained for all other uses, in any current or future media, including reprinting/republishing this material for advertising or promotional purposes, creating new collective works, for resale or redistribution to servers or lists, or reuse of any copyrighted component of this work in other works. he peer-reviewed paper is available at \url{https://ieeexplore.ieee.org/document/11202394}.\par Abdollah Masoud Darya is with SAASST and the Department of Electrical Engineering, University of Sharjah, Sharjah, UAE (email: abdollah.masoud@ieee.org). Saeed Abdallah is with the Department of Electrical Engineering, University of Sharjah, Sharjah, UAE (email: sabdallah@sharjah.ac.ae).}
}
\markboth{IEEE Transactions on Vehicular Technology,~Vol.~00, No.~0, Month~0000}%
{Shell \MakeLowercase{\textit{et al.}}: A Sample Article Using IEEEtran.cls for IEEE Journals}


\maketitle

\begin{abstract}
Massive multiple-input multiple-output low-Earth-orbit communication channels are highly time-varying due to severe Doppler shifts and propagation delays. While satellite-mobility-induced Doppler shifts can be compensated using known ephemeris data, those caused by user mobility require accurate user positioning information; the absence of such information contributes to amplified channel aging in conventional channel estimators. To address this challenge, we propose a data-aided channel estimator based on the expectation–maximization (EM) algorithm, combined with a discrete Legendre polynomial basis expansion model (DLP-BEM), to estimate the channel under imperfect Doppler compensation. The EM algorithm iteratively exploits hidden data symbols for improved channel estimation, while DLP-BEM regularizes the process by projecting the channel estimate onto a lower-dimensional subspace that mitigates estimation errors. Simulation results demonstrate the superiority of the proposed framework over existing methods in terms of normalized mean square error and symbol error rate.
\end{abstract}

\begin{IEEEkeywords}
Expectation-maximization, massive MIMO, low‐Earth‐orbit, satellite communication, basis expansion model.
\end{IEEEkeywords}

\section{Introduction}\label{intro}
Direct-to-cell (D2C) communications promise global connectivity by leveraging dense low‐Earth‐orbit (LEO) satellite constellations equipped with massive multiple-input multiple-output (mMIMO) arrays capable of highly directive beamforming \cite{andrews20246g},\cite{10844052}. However, obtaining accurate channel state information (CSI) in such systems is challenging due to long propagation delays and pronounced Doppler shifts primarily induced by satellite mobility and, to a lesser extent, by user mobility \cite{you2020massive}. While satellite-induced Doppler shifts are often mitigated via known ephemeris data, user-mobility-induced Doppler shifts necessitate precise real-time positioning through global navigation satellite system (GNSS) modules \cite{3gpp}. In practice, user Doppler compensation is hindered when users lack GNSS modules due to cost constraints, or when available GNSS signals are disrupted by jamming, or spoofing \cite{chen2025anomaly,zhu2025robust}. This leads to the \emph{imperfect Doppler compensation} scenario considered in this work, which intensifies the channel aging problem, thus degrading equalization and detection performance when pilot-based channel estimators are employed \cite{shen2022random,abdelsadek2022distributed,ying2023quasi,li2023channel}.\par
The unique characteristics of LEO satellite communications, including severe Doppler shifts, rapid channel aging, and the impracticality of dense pilot symbols due to limited spectral efficiency, require the development of estimation techniques that can reliably exploit data symbols in addition to pilots. Hence, data-aided channel estimation schemes have emerged as a robust alternative. By treating data symbols as virtual pilots, such approaches can continually track channel variations even in highly dynamic environments. For instance, data-aided variational Bayesian inference (VBI) methods have demonstrated superior performance over traditional pilot-based schemes in orthogonal time frequency space (OTFS) LEO satellite communications \cite{wang2022joint}. Similarly, decision-directed iterative least-squares formulations have been proposed to mitigate channel aging for orthogonal frequency-division multiplexing (OFDM) LEO communications \cite{darya2024semi}.\par
Due to the aging induced by the time-varying channel and accelerated by imperfect Doppler compensation, conventional decision-directed channel estimators in the literature remain limited by their reliance on hard symbol decisions, which are prone to error propagation. The iterative expectation-maximization (EM) algorithm is particularly well-suited to this setting because it leverages soft symbol decisions in each iteration, enabling more reliable use of data symbols at low signal-to-noise ratio (SNR), while its iterative nature progressively refines the channel estimate and improves robustness against Doppler-induced variations. Building on these advantages, we propose a novel data-aided channel estimator that combines EM with the discrete Legendre polynomial basis expansion model (DLP-BEM). Here, EM exploits hidden data symbols to enhance CSI estimation, whereas DLP-BEM regularizes the process by projecting the estimate onto a lower-dimensional subspace, thereby mitigating noise and alleviating the effects of imperfect Doppler compensation. We evaluate the proposed framework in terms of normalized mean square error (NMSE) and symbol error rate (SER), benchmarking its performance against both a perfectly known channel during pilot transmission and a recent data-aided approach \cite{darya2024semi}.\par
The rest of this paper is organized as follows: Section \ref{ChannelModel} presents the system model, Section \ref{ChannelEstimation} investigates the proposed EM channel estimator, Section \ref{RnD} discusses the results, and Section \ref{Conc} concludes the paper.\par

\section{System Model}\label{ChannelModel}
Consider a mMIMO LEO satellite communications system serving $K$ single-antenna user terminals (UTs) simultaneously. The satellite has a uniform planar array (UPA) consisting of $M=M_xM_y$ elements, where the numbers of $x$- and $y$-axes antenna elements are represented by $M_x$ and $M_y$, respectively. The UPA supports multi-beam transmission with full frequency reuse. The system operates in the frequency division duplexing mode and employs OFDM \cite{you2020massive}.\par
For the reasons stated in Section \ref{intro}, it is assumed that the mobile UTs do not have access to their accurate real-time position, nor their movement speed or direction. They instead have access to a coarse estimate of their position, e.g., within a few kilometers \cite{zhu2025robust}. This coarse estimate can be provided by a mMIMO satellite by leveraging its multi-beam nature to identify the origin of the signal within a cell and feed it back to the UT \cite{pat1}.\par
The uplink channel between the satellite and UT $k$, at time $t$ and frequency $f$, consists of two components: line of sight (LoS) and non-line of sight (NLoS) \cite{yue2022block}, i.e.,
\begin{equation}\label{eq1}
\boldsymbol{h}_k(t,f)=\sqrt{\beta_k}\left(h_k^{\text{LoS}}(t,f)+h_k^{\text{NLoS}}(t,f)\right)\cdot\boldsymbol{a}_{k}.
\end{equation}
The large-scale fading gain ${\beta_k}$ per UT $k$ is represented as ${\beta_k}=\left(\frac{\lambda_c}{4\pi d_k}\right)^2$, and $\lambda_c=\mathfrak{c}/f_c$, where $f_c$ is the carrier frequency, $\mathfrak{c}$ is the speed of light, and $d_k$ is the distance between the satellite and UT $k$ \cite{bjornson2024introduction}.
Furthermore, $\boldsymbol{a}_{k}$ is the UPA response vector represented by \cite{you2020massive}
\begin{equation}
\boldsymbol{a}_{k} = \boldsymbol{v}_x\left(\sin\left(\theta^y_k\right)\cos\left(\theta^x_k\right)\right) \otimes \boldsymbol{v}_y\left(\cos\left(\theta^y_k\right)\right) \in \mathbb{C}^{M\times 1}~,
\end{equation}
where $\theta^y_k$ and $\theta^x_k$ are the angles associated with the $x$- and $y$-axes of the propagation paths for UT $k$ (due to the satellite’s high altitude, all paths for a given UT share the same angles \cite{you2020massive}), respectively, and $\otimes$ represents the Kronecker product. The array vectors $\boldsymbol{v}_x$ and $\boldsymbol{v}_y$, where $d\in\left\{x,y\right\}$ and $\boldsymbol{v}_d\left(\mathcal{D}\right) \in \mathbb{C}^{M_d\times 1}$, can be represented as
\begin{equation}
\boldsymbol{v}_d\left(\mathcal{D}\right)=\frac{\left[1, \exp\left\{-j\pi\mathcal{D}\right\}, \cdots, \exp\left\{-j\pi\left(M_d-1\right)\mathcal{D}\right\}\right]^T}{\sqrt{M_d}}.
\end{equation}
The vector $\boldsymbol{a}_{k}$ can be deduced from known satellite and coarse UT positions \cite{3gpp}. Due to the high altitude of the LEO satellite, $\theta^x_k$ and $\theta^y_k$ will change very slowly, even with a highly mobile user \cite{you2024integrated}. For instance, for a satellite at an altitude of $600\,\mathrm{km}$, if a user moves by $1\,\mathrm{km}$, this translates to a change in position less than $ 0.1\degree$, as viewed from the satellite. Consequently, for a single uplink frame, it is assumed that the inaccuracy in UT $k$'s position has a negligible effect on $\boldsymbol{a}_{k}$, which remains constant during the uplink frame.\par Considering the high altitude and velocity of the LEO satellite, the uplink channel simplifies to \cite{zhang2022deep,darya2024semi}
\begin{equation}
\begin{split}
\boldsymbol{h}_k(t,f)=&\sqrt{\frac{\beta_k}{\kappa_k+1}}\cdot\exp\left\{j2\pi t\nu_{k}^{\text{SAT}}\right\}\\&\cdot\left(h_k^{\text{LoS}}(t,f)+h_k^{\text{NLoS}}(t,f)\right)\cdot\boldsymbol{a}_{k},
\end{split}
\end{equation}
where $\kappa_k$ is the Rician factor and  $\nu_k^{\text{SAT}}$ is the satellite's Doppler shift per UT $k$. Additionally, the LoS component is represented by
\begin{equation}
h_k^{\text{LoS}}(t,f)=\sqrt{\kappa_k}\cdot\exp\left\{j2\pi \left(t\nu_k^{\text{UT-LoS}}-f\tau_k^{\text{LoS}}\right)\right\},
\end{equation}
where $\tau_k^{\text{LoS}}$ is the LoS time delay and $\nu_k^{\text{UT-LoS}}$ is the LoS UT Doppler shift per UT $k$. Furthermore, the NLoS component is represented by
\begin{equation}
 h_k^{\text{NLoS}}(t,f)\!=\!\sqrt{\frac{1}{P_k}}\sum_{p=1}^{P_k}g_{k,p}\cdot\exp\!\left\{j2\pi\left(t\nu_{k,p}^{\text{UT-NLoS}}-f\tau_{k,p}^{\text{NLoS}}\right)\right\},
\end{equation}
where $P_k$ is the number of NLoS propagation paths per UT $k$, $\tau_{k,p}^{\text{NLoS}}$ is the NLoS time delay and $\nu_{k,p}^{\text{UT-NLoS}}$ is the NLoS Doppler shift per UT $k$ and path $p$. Note that $\tau_{k,p}^{\text{NLoS}}=\tau_k^{\text{LoS}}+\tau_{k,p}^{\text{MP}}$, where $\tau_{k,p}^{\text{MP}}$ is the multipath time delay per UT $k$ and path $p$. Furthermore, $\tau_k^{\text{LoS}}\gg\tau_{k,p}^{\text{MP}}$ and the delay spread is $\Delta\tau_k=\tau_k^{\text{max}}-\tau_k^{\text{min}}$, and $g_{k,p}$ is the Rayleigh fading complex Gaussian-distributed gain with zero mean and unit variance \cite{abdelsadek2022distributed}.\par

The demodulated uplink received signal over subcarrier $c$ of OFDM symbol $s$ is represented by \cite{you2020massive} 
\begin{equation}
\label{eq6}
\boldsymbol{y}_{s,c}= \sum_{k=1}^{K}\boldsymbol{h}_{k,s,c}\cdot x_{k,s,c}+\boldsymbol{z}_{s,c} \in\mathbb{C}^{M\times 1}~,
\end{equation}
where $\boldsymbol{h}_{k,s,c}=h_{k}\left(sT_{sl},cf_s\right)\cdot\boldsymbol{a}_k$, for $c=1,\dots,N_{sc}$, and $s=1,\dots,S$, where $N_{sc}$ is the number of subcarriers, $f_s$ is the subcarrier spacing, and $S$ is the total number of OFDM symbols \cite{you2024integrated}. Furthermore, $x_{k,s,c} \in \mathbb{C}$ is the symbol transmitted by UT $k$, and $\boldsymbol{z}_{s,c} \sim \mathcal{C}\mathcal{N}\left(\boldsymbol{0},\sigma^2\boldsymbol{I}\right)$ is the complex additive white Gaussian noise (AWGN) with zero mean and covariance $\sigma^2\boldsymbol{I}$. The OFDM symbol duration is $T_{sl}=N_{sc}T_s+T_{cp}$, where $T_s=(2\mathrm{B})^{-1}$ is the sampling period with system bandwidth $\mathrm{B}$. Furthermore, the cyclic prefix duration is represented by $T_{cp}=N_{cp}T_s>\Delta\tau_k$, where $N_{cp}$ is the number of cyclic prefix samples.\par

All observations at the satellite side for $S$ symbols for all $K$ UTs are collected into the matrix $\boldsymbol{Y}_c \in \mathbb{C}^{S\times M}$, where $\boldsymbol{Y}_c=\left[\boldsymbol{y}_{1,c},\ldots,\boldsymbol{y}_{S,c}\right]^T$. Additionally 
\begin{equation}\label{chaneq}
\boldsymbol{Y}_c=\left(\boldsymbol{H}_c \odot \left(\boldsymbol{X}_c \odot \boldsymbol{\Omega}_c\right)\right)^T\!\cdot\!\boldsymbol{A}+\boldsymbol{Z}_c,
\end{equation}
where $\odot$ denotes the Hadamard product, $\boldsymbol{Z}_c\in \mathbb{C}^{S\times M}$ is the AWGN matrix, $\boldsymbol{A}=\left[\boldsymbol{a}_{1},\ldots,\boldsymbol{a}_{K}\right]^T\in\mathbb{C}^{K\times M}$,  and
\begin{equation}
\begin{split}
    \boldsymbol{H}_c=& \begin{bmatrix} 
    h_{1,1,c} & \dots & h_{1,S,c}\\
    \vdots & \ddots & \vdots\\
    h_{K,1,c} & \dots & h_{K,S,c}
    \end{bmatrix}  \in \mathbb{C}^{K\times S},\\
    \boldsymbol{X}_c=& \begin{bmatrix}
    x_{1,1,c} & \dots & x_{1,S,c}\\
    \vdots & \ddots & \vdots\\
    x_{K,1,c} & \dots & x_{K,S,c}
    \end{bmatrix} \in \mathbb{C}^{K\times S}.\\
\end{split}
\end{equation}
The Doppler and delay compensation matrix is defined as
\begin{equation}
\boldsymbol{\Omega}_c=\exp\left\{-j2\pi\left( \boldsymbol{T}\otimes\boldsymbol{\mathcal{V}}_c^\text{SAT}-\frac{c}{T_{sl}}\boldsymbol{1}\otimes\boldsymbol{\mathcal{T}}^\text{LoS}\right)\right\},
\end{equation}
where $\boldsymbol{T}=\left[T_{sl}+T_{cp},\dots,S\left(T_{sl}+T_{cp}\right)\right]$. Also, $\boldsymbol{\mathcal{V}}_c^{\text{SAT}}=\left[\nu_{1}^{\text{SAT}},\dots,\nu_{K}^{\text{SAT}}\right]^T$  represents the satellite Doppler compensation and $\boldsymbol{\mathcal{T}}^\text{LoS}=\left[\tau_1^{\text{LoS}},\dots,\tau_K^{\text{LoS}}\right]^T$ represents the LoS propagation delay compensation, both performed at the UT side \cite{11143887}. Moreover, $\boldsymbol{1}$ is a row vector of ones with length $S$.\par
Due to the lack of accurate user position data, users cannot compensate for the $\nu_k^{\text{UT-LoS}}$ Doppler component. Additionally, due to user mobility, the path-dependent $\nu_{k,p}^{\text{UT-NLoS}}$ and $\tau_{k,p}^{\text{NLoS}}$ will also remain uncompensated. These uncompensated effects will accelerate the channel aging process, thus necessitating accurate channel estimators that can tackle rapidly time-varying channels while reducing the pilot overhead. Note that if $\boldsymbol{\mathcal{V}}_c^{\text{SAT}}$ is unknown or contains significant errors, it can be estimated using the methods proposed in \cite{yeh2024efficient}.\par

\section{Proposed EM Estimator}\label{ChannelEstimation}

Basis expansion models (BEMs) are widely used to approximate rapidly time-varying channels by projecting the channel onto a subspace spanned by orthonormal basis functions \cite{senol2012nondata}. This projection not only provides a compact representation of the channel dynamics but also acts as a smoothing operator, effectively suppressing high-frequency estimation noise, and is used in this work to improve the stability of subsequent EM iterations.\par
While the discrete Karhunen--Lo\`{e}ve BEM (DKL-BEM) is optimal in terms of MSE, it is highly sensitive to channel statistics variations. In contrast, the DLP-BEM offers a more robust alternative, as it is independent of channel statistics, and delivers performance that is comparable to that of the DKL-BEM \cite{senol2012nondata}.\par
The channel $h_k(s)$, for user $k$, at symbol $s$, can be represented as the weighted sum of $S$ orthonormal basis functions $\psi_q\in\mathbb{R}$ \cite{senol2012nondata}, i.e.,
\begin{equation}
    h_k(s)=\sum_{q=1}^{S}\psi_q(s)\,c_k(q),
\end{equation}
where $c_k(q)\in\mathbb{C}$ is the expansion coefficient. Additionally, $h_k(s)$ can be well approximated by $\tilde{h}_k(s)$, the weighted sum of $D$ orthonormal basis functions, where $D\ll S$, and
\begin{equation}
    \tilde{h}_k(s)=\sum_{q=1}^{D}\psi_q(s)\,c_k(q).
\end{equation}
Physically, $D$ controls the dimensionality of the subspace onto which $h_k(s)$ is projected, and thus determines how much of the channel’s temporal dynamics can be represented. By considering the orthogonal nature of the basis functions, the expansion coefficient can be found through the inverse transformation
\begin{equation}
    c_k(q)=\sum_{s=1}^{S}\psi_q(s)\,h_k(s).
\end{equation}
For UT $k$, the channel and expansion coefficients can be expressed as \cite{senol2012nondata}
\begin{equation}
\begin{split}
    \tilde{\boldsymbol{h}}_k&=\boldsymbol{\Psi}\boldsymbol{c}_k,\\
    \boldsymbol{c}_k&=\boldsymbol{\Psi}^T\boldsymbol{h}_k,
\end{split}
\end{equation}
where $\tilde{\boldsymbol{h}}_k=[\tilde{h}_k(1),\ldots,\tilde{h}_k(S)]^T\in\mathbb{C}^{S\times 1}$, $\boldsymbol{c}_k=[c_k(1),\ldots,c_k(D)]^T\in\mathbb{C}^{D\times 1}$, $\boldsymbol{\Psi}=[\boldsymbol{\psi}(1),\ldots,\boldsymbol{\psi}(S)]^T\in\mathbb{R}^{S\times D}$, and $\boldsymbol{\psi}=[\psi_1(s),\ldots,\psi_D(s)]^T$, for $s=1,2,\dots, S$. In this case ${\tilde{\boldsymbol{h}}}_k$ is the DLP-BEM projection of ${\boldsymbol{h}}_k$.\par

For $s =1,2,\dots, S$, $q =3,4,\dots,D$, the Legendre polynomials $\eta_q(s)$ and their corresponding normalization coefficients $\zeta_q$ can be computed as \cite{senol2012nondata}
\begin{equation}
\begin{split}
\eta_q(s) &= \frac{(2q - 3)(S - 2n + 1)}{(q-1)(S - q+1)} \eta_{q-1}(s)\\ &- \frac{(q - 2)(S + q - 2)}{(q-1)(S - q-1)} \eta_{q-2}(s),\\
\zeta_q &= \sqrt{\frac{(2q - 3)(S + q -1)}{(2q - 1)(S - q +1)}} \zeta_{q-1},~ \psi_q(s) = \frac{\eta_q(s)}{\zeta_q}
\end{split}
\end{equation}
where $\eta_1(s)=1$, $\eta_2(s)=1-\frac{2(s-1)}{S-1}$, $\zeta_1=\sqrt{S}$, and $\zeta_2 = \sqrt{\frac{S(S + 1)}{3(S - 1)}}$.\par
For the system in \eqref{chaneq}, the contribution of $\boldsymbol{A}$ can be removed through the pseudo-inverse $\boldsymbol{A}^{\dagger}\in \mathbb{C}^{M\times K}$, obtaining
\begin{equation}\label{eq_trans}
\tilde{\boldsymbol{Y}}_c=\boldsymbol{Y}_c\boldsymbol{A}^{\dagger}=\left(\boldsymbol{H}_c \odot \left(\boldsymbol{X}_c \odot \boldsymbol{\Omega}_c\right)\right)^T+\boldsymbol{Z}_c\boldsymbol{A}^{\dagger},
\end{equation}
where
\begin{equation}
    \mathrm{Cov}\{\boldsymbol{Z}_c\boldsymbol{A}^{\dagger}\}=\sigma^2(\boldsymbol{A}^{\dagger})^H\boldsymbol{A}^{\dagger}\approx \sigma^2 \boldsymbol{I}_{K\times K},
\end{equation}
for sufficiently large $M$, since $\boldsymbol{A}$ is asymptotically orthogonal \cite{you2020massive}. Furthermore, antenna arrays in mMIMO LEO satellites, such as D2C satellites, are quite large (more than $2\,000$ elements in some cases \cite{tuzi2023satellite}). Therefore, the effect of the transformation in \eqref{eq_trans} on the noise covariance is negligible.\par 

Let $\tilde{y}_{k,s}$ be the received signal from UT $k$ at symbol $s$, corresponding to a single element from $\tilde{\boldsymbol{Y}}_c\in \mathbb{C}^{S\times K}$. The posterior probability
\begin{equation}\label{posterior}
P(\xi_{\mathfrak{n}}|\tilde{y}_{k,s},\hat{h}_{k,s}^{(\imath)})=
\frac{\exp\left(\frac{-1}{\sigma^2}\vert\tilde{y}_{k,s}-\hat{h}_{k,s}^{(\imath)}\xi_{\mathfrak{n}}\vert^2\right)}{\sum_{\acute{\mathfrak{n}}=1}^{N_{\xi}}\exp\left(\frac{-1}{\sigma^2}\vert\tilde{y}_{k,s}-\hat{h}_{k,s}^{(\imath)}\xi_{\acute{\mathfrak{n}}}\vert^2\right)},
\end{equation}
represents the probability of the transmission of the hidden symbol $\xi_{\mathfrak{n}}$ for $\mathfrak{n}=1,\dots,N_{\xi}$, given $\tilde{y}_{k,s}$ and the current channel estimate $ \hat{h}_{k,s}^{(\imath)}$, where $N_{\xi}$ denotes the number of possible symbol hypotheses.\par
The $\imath$th iteration of the EM algorithm involves two steps. The first is the expectation step (E-step), which entails formulating the expected value of the complete-data log-likelihood function
\begin{equation}
\mathcal{L}\left(\tilde{y}_{k,s}|x_{k,s},h_{k,s}\right)\propto \vert\tilde{y}_{k,s}-{h}_{k,s}x_{k,s}\vert^2    
\end{equation}
 using \eqref{posterior}, given the observations of $\tilde{y}_{k,s}$ and the estimate $\hat{h}_{k,s}^{(\imath)}$, i.e., 
\begin{equation}\label{e-step1}
Q(h_{k,s}|\hat{h}_{k,s}^{(\imath)})=\mathbb{E}\left[\mathcal{L}\left(\tilde{y}_{k,s}|x_{k,s},h_{k,s}\right)|\tilde{y}_{k,s},\hat{h}_{k,s}^{(\imath)}\right].
\end{equation}
The second step is the maximization step (M-step). It involves maximizing \eqref{e-step1} with respect to $h_{k,s}$ to acquire the updated estimate
\begin{equation}
\hat{h}_{k,s}^{(\imath+1)}=\argmax_{h_{k,s}}~Q(h_{k,s}|\hat{h}_{k,s}^{(\imath)}).
\end{equation}
The E-step can be written as
\begin{equation}\label{e-step2}
Q(h_{k,s}|\hat{h}_{k,s}^{(\imath)})\propto\sum_{\mathfrak{n}=1}^{N_{\xi}}P(\xi_{\mathfrak{n}}|\tilde{y}_{k,s},\hat{h}_{k,s}^{(\imath)})\cdot\vert\tilde{y}_{k,s}-{h}_{k,s}\xi_{\mathfrak{n}}\vert^2.  
\end{equation}
The maximization step involves finding the derivative of \eqref{e-step2} with respect to  $h_{k,s}^*$. As a result, the M-step, after incorporating DLP-BEM, can be written as
\begin{equation}\label{eq_em}
\begin{split}
\frac{\partial Q}{\partial h_{k,s}^*} 
=&-\sum_{\mathfrak{n}=1}^{N_{\xi}}P(\xi_{\mathfrak{n}}|\tilde{y}_{k,s},\hat{h}_{k,s}^{(\imath)})
\cdot(\tilde{y}_{k,s}\xi_{\mathfrak{n}}^*-h_{k,s}\xi_{\mathfrak{n}} \xi_{\mathfrak{n}}^*)=0,\\
\hat{h}_{k,s}^{(\imath+1)}
=&\boldsymbol{\psi}(s)\boldsymbol{\psi}(s)^T
\frac{\sum_{\mathfrak{n}=1}^{N_{\xi}}P(\xi_{\mathfrak{n}}|\tilde{y}_{k,s},\hat{h}_{k,s}^{(\imath)})\cdot\tilde{y}_{k,s}\xi_{\mathfrak{n}}^*}{\sum_{\mathfrak{n}=1}^{N_{\xi}}P(\xi_{\mathfrak{n}}|\tilde{y}_{k,s},\hat{h}_{k,s}^{(\imath)})\cdot\vert\xi_{\mathfrak{n}}\vert^2}.
\end{split}
\end{equation}

The EM algorithm can be iterated over $\imath=1,\ldots,N_\text{EM}$ to improve its estimation performance. It is important to note that the EM formulation in \eqref{eq_em} does not utilize outdated pilot symbols and relies completely on the hidden data. This is crucial in rapidly time-varying channels, such as the one considered in this work, since the outdated pilot symbols would degrade EM performance.\par

The proposed EM estimator for all $K$ UTs and $S$ symbols is shown in Algorithm \ref{alg1}, where $\boldsymbol{\Xi}=\{\Xi_\mathfrak{n}\}_{\mathfrak{n}=1}^{N_\xi}\in \mathbb{C}^{K\times S \times N_\xi}$, $\abs(\cdot)$ represents the element-wise absolute operation, $\oslash$ denotes the Hadamard division, and 
\begin{equation}
\begin{split}
\Xi_\mathfrak{n}=& \begin{bmatrix} 
\xi_{1,1} & \dots & \xi_{1,S}\\
\vdots & \ddots & \vdots\\
\xi_{K,1} & \dots & \xi_{K,S}
\end{bmatrix}  \in \mathbb{C}^{K\times S}.
\end{split}
\end{equation}

A pilot-based estimator such as the least-squares (P-LS) estimator from \cite{darya2024semi}, or the minimum mean square error (MMSE) estimator from \cite{li2023channel}, can be used to obtain the initial estimate $\hat{\boldsymbol{H}}_c^{(1)}$ from a prior time slot. By incorporating the product $\boldsymbol{\Psi}\boldsymbol{\Psi}^{T}$ into the EM channel estimator, the performance of the EM method in estimating time-varying channels is enhanced. Specifically, after each M-step, the channel estimate is projected onto the subspace spanned by the first $D$ Legendre basis functions. This projection exploits the correlation between neighboring CSI values in the time domain, thereby filtering out noise-dominated high-frequency variations while retaining the channel estimate's dominant temporal dynamics. Consequently, the DLP-BEM serves as a subspace regularization mechanism that stabilizes the EM iterations and improves estimation accuracy under imperfect Doppler compensation.\par

\begin{algorithm*}[!t]
\caption{\small{Proposed EM Channel Estimator}}
\begin{algorithmic}[1]
\renewcommand{\algorithmicrequire}{\textbf{Input:}}
\renewcommand{\algorithmicensure}{\textbf{Output:}}
\REQUIRE Initial channel estimate $\hat{\boldsymbol{H}}_c^{(1)}$
\ENSURE Updated EM channel estimate $\hat{\boldsymbol{H}}_c^{\text{EM}} = \hat{\boldsymbol{H}}_c^{(N_{\text{EM}})}$
\FOR {$\imath = 1$ to $N_\text{EM}$}
\STATE 
{\scriptsize
$ \displaystyle
\begin{aligned}
\boldsymbol{\Gamma}({\Xi}_{\mathfrak{n}} | \tilde{\boldsymbol{Y}}_c, \hat{\boldsymbol{H}}_c^{(\imath)}) =
\left({
\exp\left\{\frac{-1}{\sigma^2} \abs\left(\tilde{\boldsymbol{Y}}_c - \hat{\boldsymbol{H}}_c^{(\imath)} \odot {\Xi}_{\mathfrak{n}}\right)^2 \right\}
}\right)\oslash\left({
\sum_{\acute{\mathfrak{n}}=1}^{N_{\xi}} \exp\left\{\frac{-1}{\sigma^2} \abs\left(\tilde{\boldsymbol{Y}}_c - \hat{\boldsymbol{H}}_c^{(\imath)} \odot {\Xi}_{\acute{\mathfrak{n}}} \right)^2 \right\}
}\right)
\end{aligned}$}
\STATE 
{\scriptsize
$ \displaystyle
\begin{aligned}
\hat{\boldsymbol{H}}_c^{(\imath+1)} = \boldsymbol{\Psi} \boldsymbol{\Psi}^{T}\left({\sum_{\mathfrak{n}=1}^{N_{\xi}} \boldsymbol{\Gamma}({\Xi}_{\mathfrak{n}}|\tilde{\boldsymbol{Y}}_c, \hat{\boldsymbol{H}}_c^{(\imath)}) \cdot (\tilde{\boldsymbol{Y}}_c {\Xi}_{\mathfrak{n}}^*)}\right)\oslash\left({\sum_{{\mathfrak{n}}=1}^{N_{\xi}}\boldsymbol{\Gamma}({\Xi}_{\mathfrak{n}}|\tilde{\boldsymbol{Y}}_c, \hat{\boldsymbol{H}}_c^{(\imath)}) \cdot \abs\left({\Xi}_{\mathfrak{n}}\right)^2}\right)
\end{aligned}$}
\ENDFOR
\end{algorithmic}
\label{alg1}
\end{algorithm*}

The complexity of the proposed EM estimator can be defined as follows. The $\boldsymbol{A}^{\dagger}$ operation has a complexity of $\mathcal{O}\left(MK^2\right)$ and is only recomputed when satellite or UT positions change significantly \cite{you2020massive}. The complexity of the product $\boldsymbol{Y}_c\boldsymbol{A}^{\dagger}$ is $\mathcal{O}\left(MKS\right)$, the element-wise multiplication and division in the exponential has a complexity of $\mathcal{O}\left(KS\right)$, and the summation over $\boldsymbol{\Xi}$ introduces a complexity of $\mathcal{O}\left(N_{\xi}KS\right)$. Additionally, the product $\boldsymbol{\Psi}\boldsymbol{\Psi}^T$ has the complexity $\mathcal{O}\left(K^2D\right)$, and the multiplication of this product by the summation over $\boldsymbol{\Xi}$ has the complexity $\mathcal{O}\left(K^2S\right)$. Therefore, the total complexity per EM iteration $\imath$ is $\mathcal{O}\left(N_{\xi}KS+K^2(D+S)+MKS\right)$. Since $\boldsymbol{A}^{\dagger}$ and $\boldsymbol{\Psi}\boldsymbol{\Psi}^T$ are precomputed offline, the run-time complexity per EM iteration $\imath$ is $\mathcal{O}\left(KS(N_{\xi}+K+M)\right)$.\par

\section{Results and Discussion}\label{RnD}
This work considers the reference parameters from 3GPP \cite{3GPP38.821,3gpp}, as listed in Table \ref{table1}.
\begin{table}[!t]
\caption{\small{Channel parameters\label{table1}}}
\centering
\begin{tabular}{|l|c|}
\hline
\textbf{Parameter} & \textbf{Value} \\
\hline
Carrier Frequency $f_c$ & $2\,\mathrm{GHz}$\\
\hline
Bandwidth $B$ & $15.36\,\mathrm{MHz}$\\
\hline
Subcarrier Spacing $f_s$ & $60\,\mathrm{kHz}$\\
\hline
Number of Subcarriers $N_{sc}$ & $256$\\
\hline
Number of UTs $K$ & $10$\\
\hline
Antenna Array Size $M_xM_y$ & $16\times16$\\
\hline
Satellite Altitude & $600\,\mathrm{km}$\\
\hline
Maximum Satellite Doppler Shift & $\pm48\,\mathrm{kHz}$\\
\hline
Minimum Elevation Angle & $30\degree$\\
\hline
Delay Spread $\Delta\tau_k$ & $250\,\mathrm{ns}$\\
\hline
Rician Factor $\kappa_k$ & $10\,\mathrm{dB}$\\
\hline
Maximum UT Speed & $160\,\mathrm{km/h}$\\
\hline
Maximum Number of Propagation Paths & $5$\\
\hline
Number of pilot-symbols per uplink frame & $5$\\
\hline
Number of data-symbols per uplink frame & $50$\\
\hline
Modulation of data-symbols & $16$-QAM\\
\hline
\end{tabular}
\end{table}
The reference time-varying channel
\begin{equation}
\begin{split}
&\boldsymbol{h}_{s}^\text{ref}=\bigl[h_{1,s}\cdot\exp\left\{-j2\pi\left(t\nu_{1}^{\text{SAT}}-f\tau_1^{\text{LoS}}\right)\right\},\dots,\\&h_{K,s}\cdot\exp\left\{-j2\pi\left(t\nu_{K}^{\text{SAT}}-f\tau_K^{\text{LoS}}\right)\right\}\bigr]^T\in\mathbb{C}^{K\times 1}
\end{split}
\end{equation}
is used to calculate the NMSE of the estimate $\boldsymbol{\hat{H}}$ using
\begin{equation}\label{NMSE}
\text{NMSE}=\frac{\left\Vert\vect\left(\left[\boldsymbol{h}_{s}^\text{ref},\dots,\boldsymbol{h}_{S}^\text{ref}\right]\right)-\vect\left(\boldsymbol{\hat{H}}\right)\right\Vert^2}{\left\Vert\vect\left(\left[\boldsymbol{h}_{s}^\text{ref},\dots,\boldsymbol{h}_{S}^\text{ref}\right]\right)\right\Vert^2},
\end{equation}
where $\vect(\cdot)$ represents the vectorization operation, and the SNR is represented as
\begin{equation}
\text{SNR}\!=\!10\log_{10}\frac{\!\left\Vert\vect\left(\left(\left[\boldsymbol{h}_{1}^\text{ref},\dots,\boldsymbol{h}_{S}^\text{ref}\right] \odot\boldsymbol{X}\right)^T\cdot\boldsymbol{A}\right)\right\Vert^2}{\left(SM\right)\sigma^2}.
\end{equation}

What follows is a comparison of the NMSE and SER performances of the 1) proposed EM estimator, 2) P-LS estimator from \cite{darya2024semi} used as an initial EM estimate, 3) data-aided decision-directed least-squares (MDDLS) estimator from \cite{darya2024semi} combined with DLP-BEM, and 4) an optimal pilot-based estimator (PB) represented by perfect channel knowledge during the pilot transmission phase. It is assumed that in a single uplink frame five Zadoff--Chu-based pilot symbols \cite{li2023channel} are transmitted, followed by $50$ data symbols, i.e., $S=55$. The NMSE was calculated over the data transmission phase $s=6,\dots,55$, where $s=1,\dots,5$ represents the training phase. For the PB and the P-LS estimators, the average channel over the training phase was considered. This step ensures that the comparison is fairer since without averaging the PB and P-LS estimates would be more prone to channel aging. Note that $N_\text{EM}=10$, unless otherwise specified.\par

\begin{figure}[!t]
\centering
\includegraphics[width=1\linewidth]{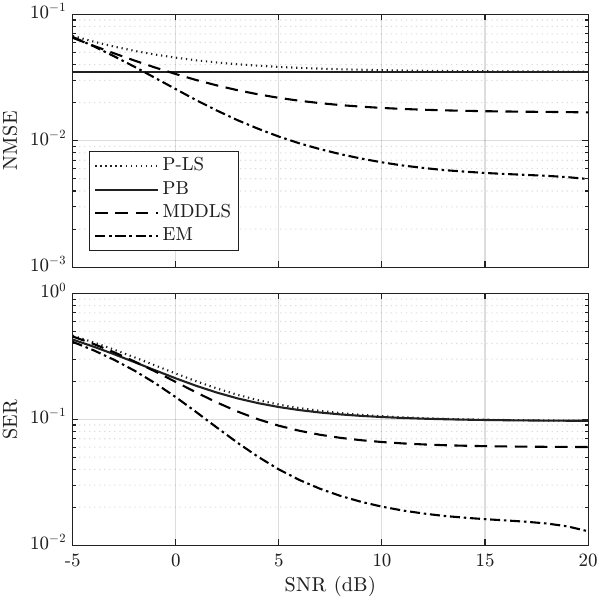}
\caption{\small{NMSE and SER vs. SNR of the proposed EM and benchmark estimators.}}
\label{Fig_1}
\end{figure}

As shown in Fig. \ref{Fig_1}, the proposed EM estimator outperforms all other estimators in terms of NMSE for SNR $>-2\,\mathrm{dB}$. The superior performance of the proposed EM estimator at low SNR can be attributed to its utilization of soft decisions compared to the hard decisions made by the MDDLS and P-LS estimators, in addition to its use of noise statistics and incorporation of the DLP-BEM. At high SNR, EM's iterative nature allows it to outperform other estimators. Additionally, the converging performance of the P-LS estimator to the PB at high SNR proves the suitability of P-LS as an initial estimate for the EM estimator.\par 

The superior estimation performance of the EM estimator is reflected in its lower SER, as illustrated in Fig. \ref{Fig_1}. This SER value was obtained by using the channel estimate provided by each respective estimator with an MMSE equalizer, followed by a minimum distance detector \cite{jiang2011performance}. The proposed EM estimator shows superior SER performance compared to other approaches for all SNR values. Specifically, the EM estimator achieves an SER of $10^{-1}$ at an SNR of $1.5\,\mathrm{dB}$, corresponding to an SNR gain of approximately $2.5\,\mathrm{dB}$ over the MDDLS estimator, and more than $11\,\mathrm{dB}$ over both the PB and P-LS estimators. These findings demonstrate that significant improvements in SER can be attained by employing the proposed EM approach in combination with the DLP-BEM.\par

\begin{figure}[!t]
\centering
\includegraphics[width=1\linewidth]{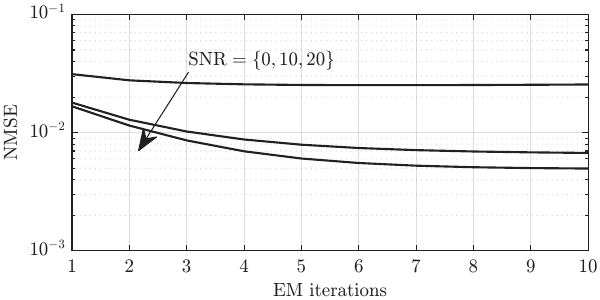}
\caption{\small{NMSE vs. number of EM iterations $N_\text{EM}$.}}
\label{Fig_2}
\end{figure}

The effect of EM iterations on the NMSE is shown in Fig. \ref{Fig_2}. The performance of the EM estimator can be seen to improve with iterations. The iterative gain in performance also increases at higher SNR. To improve the computational efficiency of the estimator, the number of iterations can be adaptively chosen based on the SNR. Thus, at low SNR, the number of EM iterations would be lower.\par

\begin{figure}[!t]
\centering
\includegraphics[width=1\linewidth]{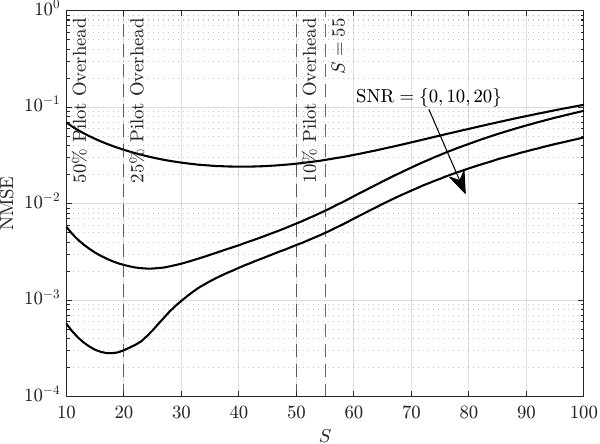}
\caption{\small{NMSE vs. uplink frame size $S$ for the proposed EM estimator.}}
\label{Fig_3}
\end{figure}

Fig. \ref{Fig_3} shows the NMSE performance of the EM estimator with respect to the uplink frame size $S$. The number of pilot symbols is fixed at $5$, while the number of data symbols is varied from $5$ to $95$. The performance of the EM estimator varies depending on the SNR. At higher SNR, shorter frames maximize EM's performance as longer frames are more affected by the smoothing process applied by the DLP-BEM. This effect is less pronounced at low SNR, where longer frames help mitigate the impact of AWGN. Since $S=55$ in this work, the pilot-overhead is $<10\%$, which is a typical value for time-varying channels \cite{qin2018time}. In practical systems, $S$ can be adapted according to the SNR to balance estimation accuracy and pilot-overhead.

\begin{figure}[!t]
\centering
\includegraphics[width=1\linewidth]{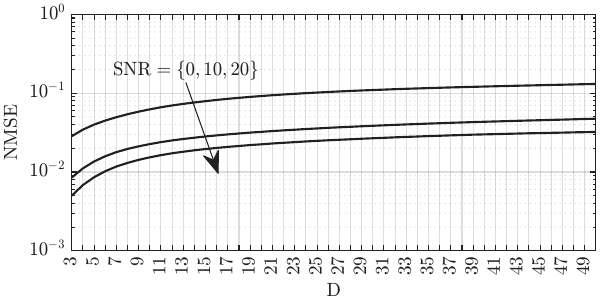}
\caption{\small{NMSE vs. $D$ for the EM estimator.}}
\label{Fig_4}
\end{figure}

Fig. \ref{Fig_4} shows the results of an exhaustive search for the optimal value of $D$. The search minimizes \eqref{NMSE} by searching from the minimum possible value of $D=3$ to the maximum possible value, which in this case is $D=50$, corresponding to the number of data symbols. The optimal value can be seen as $D=3$. Since $D$ is the minimum possible value, this also means that the complexity of the product $\boldsymbol{\Psi}\boldsymbol{\Psi}^T$ is minimized.\par

Table \ref{table2} compares the complexity of the proposed EM estimator with that of the P-LS and MDDLS estimators from \cite{darya2024semi}. The table also reports the median execution time of each estimator, computed over $10^5$ Monte Carlo iterations. The code was executed on MATLAB 2024b running on a machine equipped with an Intel Xeon 6226R CPU. It should be noted that the MDDLS used in this work differs slightly from that in \cite{darya2024semi} since the MDDLS estimate in this work has been regularized with DLP-BEM, similar to the proposed EM estimator. The median execution time of the EM estimator per iteration $\imath$ is approximately three times that of the MDDLS estimator. Yet this increase in execution time is reasonable given the improved estimation performance demonstrated in Fig. \ref{Fig_1}.  

\begin{table}[!t]
\caption{\small{Estimator Complexity and Execution Time\label{table2}}}
\centering
{\begin{tabular}{|l|c|c|}
\hline
\textbf{Estimator} & \textbf{Run-time Complexity} & \textbf{Median Execution Time}\\
\hline
P-LS & $\mathcal{O}(KS(1+M))$ & $\phantom{0}13.9\,\micro\mathrm{s}$\\
\hline
MDDLS & $\mathcal{O}(KS(K+M))$ & $\phantom{0}61.7\,\micro\mathrm{s}$\\
\hline
EM (per $\imath$) & $\mathcal{O}(KS(N_{\xi}+K+M))$ & $188.2\,\micro\mathrm{s}$\\
\hline
\end{tabular}}
\end{table} 

\section{Conclusion}\label{Conc}
This work proposed a data-aided EM-based estimator combined with DLP-BEM to estimate mMIMO LEO SATCOM channels under imperfect Doppler compensation. Owing to its iterative nature and the regularization provided by DLP-BEM, the proposed EM estimator achieved superior performance in terms of NMSE and SER compared to existing work. These results demonstrate that the proposed channel estimator is an effective solution for scenarios in which users are unable to perform perfect Doppler compensation. Future work will investigate the impact of errors in satellite Doppler compensation on the performance of channel estimators, as well as downlink channel estimation in distributed mMIMO SATCOM systems where multiple satellites jointly serve a single user.

\bibliographystyle{IEEEtran}
\bibliography{main.bib}

\end{document}